\documentclass{appolb}
\usepackage{graphicx,amsmath,amsthm,amssymb,cite,enumerate}
\usepackage{authblk} 
\usepackage[usenames]{color}

\def \d {{\rm d}}
\def \bF {\mbox{\boldmath{$F$}}}

\def \bk {\mbox{\boldmath{$k$}}}

\newcommand*\bA{\ensuremath{\boldsymbol{A}}}

\newcommand{\be}{\begin{equation}}
\newcommand{\ee}{\end{equation}}

\newcommand{\beq}{\begin{eqnarray}}
\newcommand{\eeq}{\end{eqnarray}}
\newcommand{\pa}{\partial}

\newcommand{\ba}{\begin{array}}
\newcommand{\ea}{\end{array}}

\def\R{{\cal R}}
\def\F{{\cal F}}
\def\L{{\cal L}}
\def\F2{{F}}

\def\e{e}

\begin{document}
 \eqsec  
\title{Robinson-Trautman spacetimes coupled to conformally invariant electrodynamics in higher dimensions
\thanks{Presented at the 7th Conference of the Polish Society on Relativity, \L\'od\'z, Poland, 20--23.09.2021}
}

\author[1,2]{David Koko\v ska\thanks{david.kokoska@matfyz.cz}}
\author[2]{Marcello Ortaggio\thanks{ortaggio(at)math(dot)cas(dot)cz}}

\affil[1]{Institute of Theoretical Physics, Faculty of Mathematics and Physics, \newline
 Charles University, V Hole\v{s}ovi\v{c}k\'{a}ch 2, 180 00 Prague 8, Czech Republic}
\affil[2]{Institute of Mathematics of the Czech Academy of Sciences, \newline \v Zitn\' a 25, 115 67 Prague 1, Czech Republic}

\maketitle
\begin{abstract}
We summarize recent results on $D$-dimensional Robinson-Trautman solutions of Einstein's gravity in the presence of a conformally invariant non-linear electromagnetic field and a cosmological constant. These spacetimes contain static dyonic black holes with various horizon geometries and their time-dependent radiating generalizations, as well as a class of stealth solutions. Extensions to $f(R)$ and Gauss-Bonnet gravity are mentioned.
\end{abstract}
  
	
\section{Introduction}

Robinson-Trautman (RT) spacetimes \cite{RobTra60,RobTra62} are characterized by the existence of an expanding, shearfree and twistfree congruence of null geodesics, and provide a natural arena for the study of static black holes and their time-dependent generalizations, as well as other radiative spacetimes (see the reviews \cite{Stephanibook,GriPodbook} and references therein). This class of metrics can be defined in arbitrary dimension $D$ \cite{RobTra83,PodOrt06}, and $D>4$ Einstein-Maxwell solutions have been studied in \cite{PodOrt06,OrtPodZof08,OrtPodZof15}. 

While the linear Maxwell theory is not conformally invariant when $D\neq4$, a conformally invariant non-linear electrodynamics in $D$ dimensions has been proposed in
\cite{HasMar07}. Considering a minimal coupling to Einstein gravity, the action is given by
\be
  S=\int \d^Dx\sqrt{-g}\left[\frac{1}{\kappa} \left( R - 2 \Lambda \right)-2\beta\F2^{D/4}\right] , \qquad \F2\equiv F_{\mu\nu}F^{\mu\nu}
	\label{action}
\ee
where $\kappa$ and $\beta$ are coupling constants, and $\bF=\d\bA$.

The corresponding equations of motion read
\beq
 & & \frac{1}{\kappa}\left(G_{\mu\nu}+\Lambda g_{\mu\nu}\right)=\beta F^{D/4-1}\left(DF_{\mu\rho}F_\nu^{\phantom{\nu}\rho}-g_{\mu\nu}F\right) , \label{Einst} \\
 & & \frac{1}{\sqrt{-g}}\pa_{\mu}\left(\sqrt{-g}F^{\frac{D}{4}-1}F^{\mu \nu} \right)=0 . \label{Max1}
\eeq

The RHS of \eqref{Einst} is traceless, so that the Ricci scalar is a constant proportional to $\Lambda$.

Let us observe that for $D\neq4$, solutions of the theory~\eqref{action} are {\em stealth} precisely when $F=0$, which also ensures that eq.~\eqref{Max1} is identically satisfied. In other words, any closed 2-form $\bF$ provides a solution to the theory~\eqref{action} in any Einstein spacetime. For $D=4$, eq.~\eqref{action} reduces to the Einstein-Maxwell action.

Since $F^\frac{D}{4}$ must be a real quantity, $D$ must be a multiple of 4 when $F<0$. Requiring the energy density to be non-negative (weak energy condition WEC) means $\beta F^{D/4-1}\ge0$, so that in the following we will assume: i) $\beta>0$ if $F>0$, or if $F<0$ with $D/4$ being odd; (ii) $\beta<0$ if $F<0$ with $D/4$ being even; (iii) $\beta$ can have any sign in the stealth case $F=0$. Since $T_{\mu\nu}$ is traceless, the strong energy condition becomes equivalent to the WEC and is thus also satisfied.

In our recent work \cite{KokOrt22} we studied the class of RT solutions to the theory~\eqref{action} under the assumption that the RT null vector field $\bk$ is an eigenvector of the electromagnetic field $\bF$ (i.e., it is aligned). In the following we summarize the main results obtained for non-stealth fields (some comments on the stealth case can be found in \cite{KokOrt22}).

\section{Static black holes}

\label{sec_static}

The RT solutions of \cite{KokOrt22} contain, in particular, a family of dyonic black holes. The metric reads
\be
  \d s^2=r^{2}h_{ij}\d x^i\d x^j{-}2\,\d u\d r-2H\d u^2 , 
	\label{ds_generic}
\ee
\be
 2H=K-\lambda r^2 - \frac{\mu}{r^{D-3}} + \frac{Q^2}{r^{D-2}} ,
 \label{Hstat}
\ee
and the electromagnetic field is
\be
 \bF = \dfrac{ e}{r^2}\d r \wedge\d u + \dfrac{1}{2}F_{ij}(x)\d x^i \wedge \d x^j ,
 \label{Fstat}
\ee
with
\be
	Q^2\equiv { 2\kappa \beta} F_0^{\frac{D}{4}-1}\left(\frac{b^2}{D-2} + e^2\right) , \quad F_0\equiv  b^2-2 e^2 , \quad  b^2\equiv F_{ik} F_{jl} h^{ij} h^{kl} . \label{F_spat}
\ee

In the above expressions, $K=0,\pm1$, $\lambda=\frac{2\Lambda}{(D-2)(D-1)}$, $\mu$, $e$ and $b$ are constants, Latin indices $i,j,\ldots=1,\ldots,D-2$ label the spatial coordinates $x^i$ (also denoted collectively simply as $x$), and the base space metric $h_{ij}(x)$ represents a Riemannian Einstein space of dimension $D-2$ and scalar curvature $\R=K(D-2)(D-3)$, with $h\equiv \det h_{ij}$.

The spatial part of $\bF$ and the base space metric must obey the following conditions
\be
 \Big(\sqrt{h}  h^{ik} h^{jl} F_{kl} \Big)_{,j}=0 , \quad F_{[ij,k]}= 0 , \quad  b^2h_{ij}=(D-2)F_{ik}F_{jl}h^{lk} .
 \label{eqs_BHs}
\ee
This implies that, when $F_{ij}\neq0$, the base space must be {\em almost-K\"ahler} \cite{Yanobook_complex} (in addition to being Einstein) and, in particular, it cannot be a round sphere (but it can be, e.g., flat, cf.  \cite{OrtPodZof08,KokOrt22}). This means that dyonic (or purely magnetic) solutions cannot be asymptotically flat. However, in the purely electric case ($F_{ij}=0$) the base manifold can be any Einstein space, and asymptotically flat solutions (with $\Lambda=0$) have been known for some time \cite{HasMar07}.

The above spacetimes generically represents black holes, which are static in regions where $H>0$. There is a timelike curvature singularity at $r=0$ and, in the region $r>0$, positive values of $r$ for which $H=0$ represent Killing horizons. Similarly as for the four-dimensional (A)dS-Reissner-Nordstr{\"o}m metrics, the structure and number of horizons depend on the signs of the parameters $\Lambda$, $K$ and $\mu$, and are essentially $D$-independent -- cf. \cite{KokOrt22} for details and plots of $H(r)$ for various values of the parameters.

Interestingly, thanks to the fact that the Ricci scalar of \eqref{Hstat} is constant, the above black hole solutions can be easily extended to theories for which the Einstein term in 
\eqref{action} is replaced by a generic $f(R)$ scalar -- examples have been given in \cite{Sheykhi12,HerOrt20_2} (see also \cite{EirFig22}).\footnote{For the same reason, metrics~\eqref{ds_generic}, \eqref{Hstat} can also be interpreted as vacuum solutions of certain $f(R)$ gravities (in which case $Q^2$ is an integration constant), cf. \cite{Hendi10,HenEslMou12,HerOrt20}.} Extensions to Gauss-Bonnet gravity are also known, for which the metric function $H$ is instead modified in a non-trivial way \cite{Hendi09,HerOrt20_2}.

\section{General (non-stealth) solution}

The family of black holes described in section~\ref{sec_static} is a subset of more general RT solutions of~\eqref{action}. The complete RT class is still described by metric~\eqref{ds_generic}, but now the electromagnetic field is
\be
 \bF = \dfrac{ e}{r^2}\d r \wedge\d u + \left( \dfrac{ e_{,i}}{r}-\xi_i \right)\d u \wedge \d x^i + \dfrac{1}{2}F_{ij}\d x^i \wedge \d x^j ,
 \label{F_summ}
\ee
where $\xi_i=\xi_i(u,x)$, and $H$ is defined by
\be
 2H=K+\frac{2}{D-2}\big(\ln \sqrt{h}\big)_{,u}r -\lambda r^2 - \frac{\mu}{r^{D-3}} + \frac{Q^2}{r^{D-2}} , \label{H} \\
\ee
where \eqref{F_spat} still applies. $K$ and $\lambda$ are as in section~\ref{sec_static}, whereas here the Einstein metric $h_{ij}=h^{1/(D-2)}(u,x)\gamma_{ij}(x)$ and the quantities $e$, $F_{ij}$, $\mu$ and $b$ in general depend on $(u,x)$. 

In addition to the second and third equations of~\eqref{eqs_BHs}, one now needs to solve the more complicated set  (where \eqref{grad_alpha} replaces the first of \eqref{eqs_BHs})
\beq 
 & & F_{ij,u}=\xi_{i,j}-\xi_{j,i} , \qquad 
\Big(F_0^{\frac{D}{4}-1}\sqrt{h}h^{ij} \xi_j \Big)_{,i} =  \Big(F_0^{\frac{D}{4}-1} \sqrt{h}   e\Big)_{,u} , \label{Fij,u} \\ 
 & & F_0^{\frac{D}{4}-1}\sqrt{h}h^{ij} e_{,j}  = \Big(F_0^{\frac{D}{4}-1} \sqrt{h}  h^{ik} h^{jl} F_{kl} \Big)_{,j} , \label{grad_alpha} \\
 & & \mu_{,i} =  {2\kappa \beta} D F_0^{\frac{D}{4}-1}\big(  e \xi_i - F_{ik} \xi_j h^{kj} \big) , \label{mu_i} \\
 & & (D-2)\mu_{,u} = -(D-1) \mu \big( \ln \sqrt{h} \big)_{,u}-  {2\kappa D \beta}  F_0^{\frac{D}{4}-1}h^{ij}\xi_i \xi_j  . \label{mu_u}
\eeq

Here the base space is {\em almost-Hermitian}. Note that eq.~\eqref{mu_u} must be modified in the special case $D=4$ \cite{RobTra62,Stephanibook,OrtPodZof08,GriPodbook,KokOrt22}. 

The line-element of the complete RT class is thus in general time-dependent. Furthermore, apart from the electric and magnetic components $e$ and $F_{ij}$, the electromagnetic field \eqref{F_summ} may contain also a radiative null term $F_{ui}$, which is related to a possible mass loss (or gain) as the retarded time $u$ evolves, encapsulated in eq.~\eqref{mu_u}. 
The energy flux along the RT null vector field $\bk= \pa_r$ is given by $\beta D F_0^{D/4-1}h^{ij}\xi_i\xi_jr^{2-D}$ \cite{KokOrt22}. 

For the sake of definiteness, a simple explicit time-dependent solution (not presented in \cite{KokOrt22}) with a flat base space $h_{ij}=\delta_{ij}$ is given for $D=8$ by
\beq
 & & F_{ij}=0 , \qquad e=e_0({c_1} u+1)^{1/5}  , \qquad \xi_i\d x^i=\frac{3 {c_1} {e_0} {x_1}+5 {c_0}}{5 \left({c_1} u+1\right)^{4/5}}\d x^1 , \nonumber \\
 & & \mu=\frac{16e_0^2 \beta  \kappa}{3c_1}\left[-\frac{(3 {c_1} {e_0} {x_1}+5 {c_0})^2}{5 ({c_1} u+1)^{1/5}}+5{c_0}^2\right]+\mu_0 \qquad (D=8) ,
\eeq
where $e_0$, $c_0$, $c_1$ and $\mu_0$ are constants. Note that the limit $c_1\to0$ gives rise to a function $\mu$ linear in $u$ (and to a static solution if one sets, further, $c_0=0$).

Equations defining marginally trapped surfaces \cite{Penrose73} and dynamical horizons \cite{AshKri02} in the above spacetimes have been also obtained in \cite{KokOrt22}. Those are to be understood as preliminary results needed in order to define an analog of the past horizon in RT spacetimes, in the spirit of \cite{Tod89}. See also \cite{LunCho94,ChoLun99,NatTaf08,PodSvi09,Svitek11,TahSvi16} for related results.

\section*{Acknowledgments}

This work has been supported by research plan RVO: 67985840 and research grant GA\v CR 19-09659S.


\end{document}